\documentclass[a4paper]{jpconf}

\usepackage{graphicx}

\def\lesssim{\mathrel{\hbox{\lower1ex\hbox{\rlap{$\sim$}\raise1ex\hbox{$<$}}}}}

\begin{document}

\title{New insights for the description of magnetic correlations inferred from $\mu$SR}

\author{P. Dalmas de R\'eotier$^{1,2}$, A. Yaouanc$^{1,2}$, A. Maisuradze$^{3,4,5}$}

\address{$^1$Univ. Grenoble Alpes, INAC-SPSMS, F-38000 Grenoble, France}
\address{$^2$CEA, INAC-SPSMS, F-38000 Grenoble, France}
\address{$^3$Department of Physics, Tbilisi State University, Chavchavadze 3, GE-0128
Tbilisi, Georgia}
\address{$^4$Physik-Institut der Universit\"at Z\"urich, Z\"urich, Switzerland}
\address{$^5$Laboratory for Muon-Spin Spectroscopy, Paul Scherrer Institute,
CH-5232 Villigen-PSI, Switzerland}

\begin{abstract}
Whenever a compound exhibits a spontaneous $\mu$SR oscillation, long-range magnetic ordering is 
usually inferred. Here we show that some caution is required.  
The coherence length needs not to be large for a spontaneous muon spin precession to be observed.
Establishing the incommensurate nature of a magnetic 
structure, solely based on $\mu$SR measurements, may not be reliable. The absence of a spontaneous muon 
precession at low 
temperature does not mean that the system under investigation does not display long-range magnetic 
ordering.

The relaxation measured in zero and longitudinal field in the quasi-static limit is usually analyzed 
in the framework of the strong-collision model, the static polarization function being taken to be 
the famous Kubo-Toyabe function. This might not be satisfactory if short-range correlation effects 
are strong. Here we propose a method based on the maximum entropy concept and reverse Monte Carlo technique
which gives results consistent with those obtained in 2013 by analytical means for the considered example.

\end{abstract}

\section{Introduction}
\label{intro}

In this report we consider problems relevant for the analysis of experimental positive 
muon spectroscopy ($\mu$SR) data recorded for a magnetic compound. First, we discuss the meaning of an observed 
spontaneous muon precession in relation to the possible long-range nature of the magnetic order. 
Using published examples, we point out the difficulty of reliably establishing solely with $\mu$SR the 
incommensurate nature of the magnetic structure. In Sec.~\ref{correlations} we present 
a numerical method for the determination of the field distribution at the muon site for a compound 
characterized by quasi-static spin dynamics. As an example, we apply the method for the quantum 
spin-ice system Yb$_2$Ti$_2$O$_7$. A discussion of our results is presented in Sec.~\ref{discussion}.
An appendix details the reverse Monte Carlo algorithm.

\section{Long- versus short-range magnetic order}
\label{long_range}

\begin{figure}
\begin{minipage}[t]{3in}
\includegraphics[width=\textwidth]{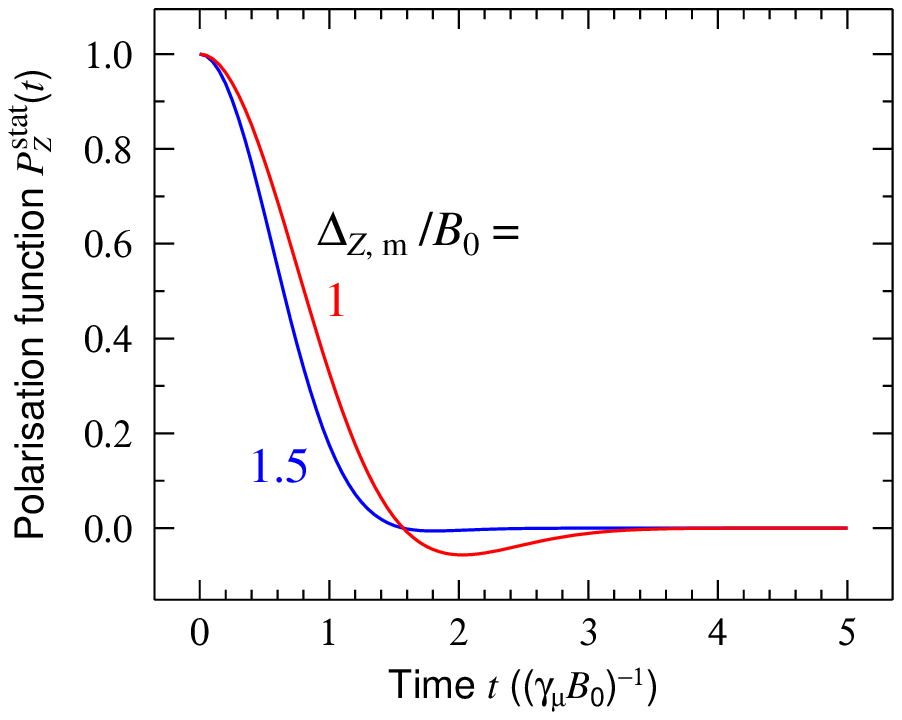}
\caption{The polarization function for two values of the ratio of the standard deviation 
of the field at the muon site over the intensity of the spontaneous field; from Ref.~\cite{Yaouanc11} by permission of Oxford University Press.
} 
\label{Book_polarisation}
\end{minipage}
\hfill
\begin{minipage}[t]{3in}
\includegraphics[width=\textwidth]{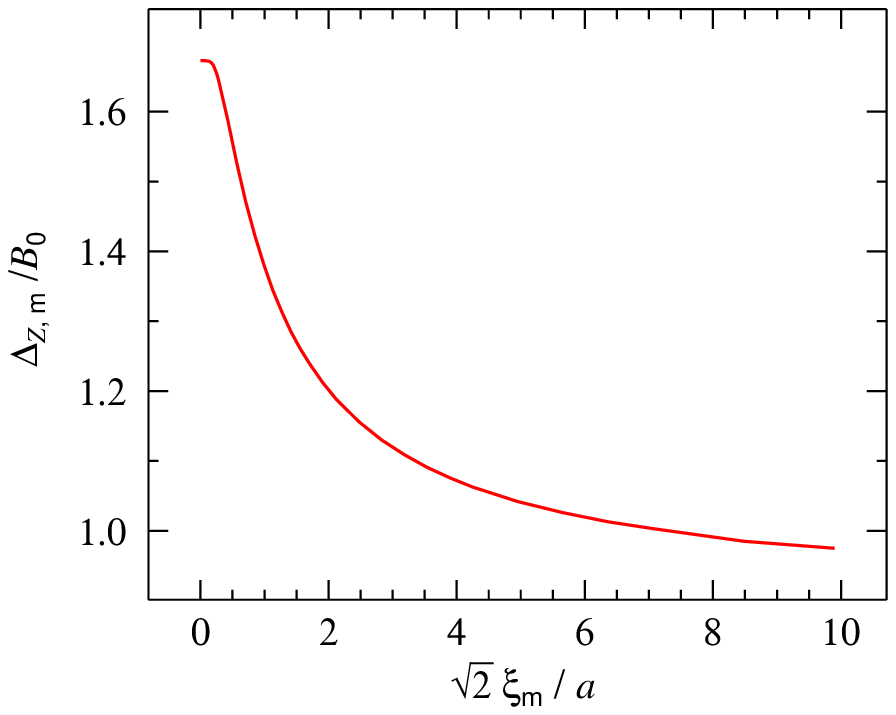}
\caption{Ratio of the standard deviation of the field at the muon site over the intensity
of the spontaneous field versus the correlation length; from Ref.~\cite{Yaouanc11} by permission of Oxford University Press.
}
\label{Book_ratio}
\end{minipage}
\end{figure}

\begin{figure}
\begin{minipage}[t]{3in}
\includegraphics[height=2.1in]{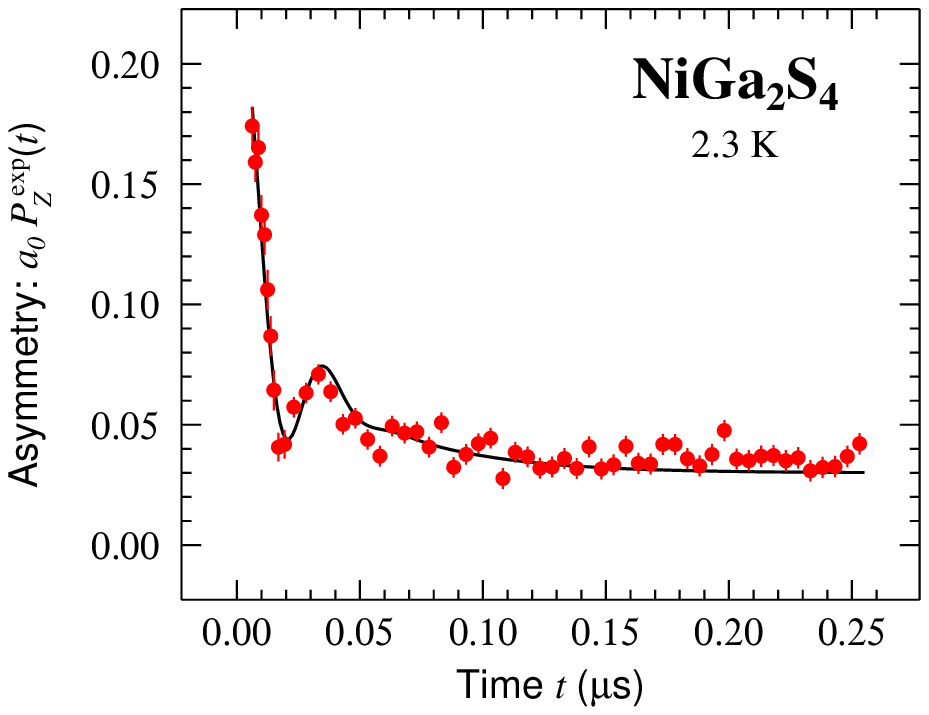}
\caption{A zero-field $\mu$SR spectrum for a powder of NiGa$_2$S$_4$ 
taken at $2.3$~K; adapted from  Ref.~\cite{Yaouanc08}. The solid line results from a fit.
} 
\label{Exp_muon}
\end{minipage}
\hfill
\begin{minipage}[t]{3in}
\includegraphics[height=2.1in]{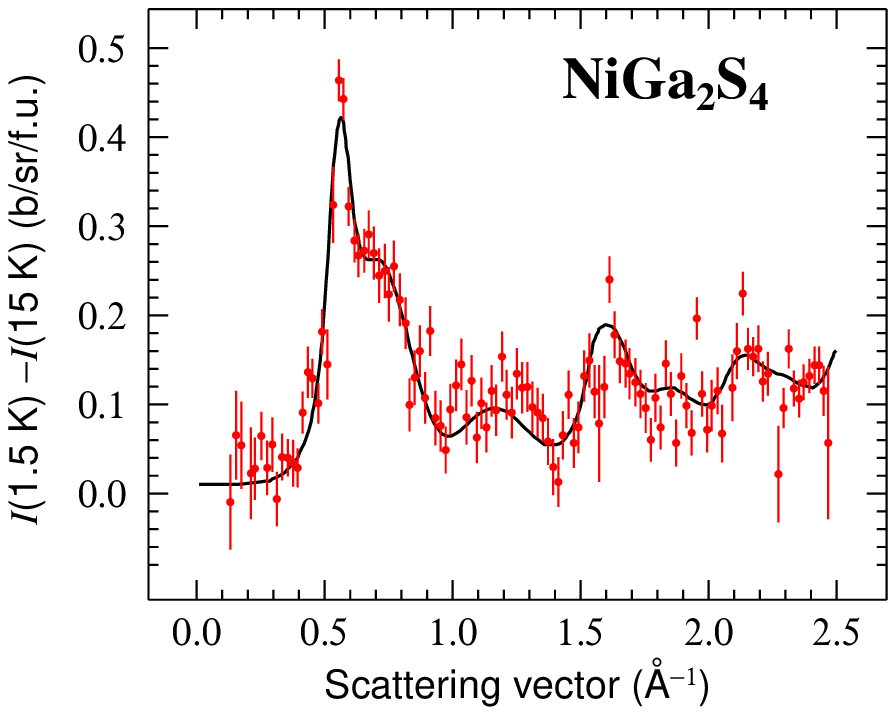}
\caption{A magnetic neutron powder diffraction pattern for NiGa$_2$S$_4$ resulting from the difference 
between the patterns recorded at $1.5$ and $15$~K. The magnetic reflections are quite broad. 
They correspond to a correlation length $\xi_{\rm m} = 2.5 \, (3)$~nm. The solid line results 
from a fit. Adapted from Ref.~\cite{Nakatsuji05}. }
\label{Exp_neutron}
\end{minipage}
\end{figure}

A large fraction of the $\mu$SR studies are performed with the purpose of 
determining whether a magnetic material exhibits a magnetic transition. If a spontaneous $\mu$SR 
oscillation is detected, the material is usually proposed to be characterized by a long-range 
magnetic order. This may not always be justified as it was shown theoretically in 
Ref.~\cite{Yaouanc11}. Let us consider the polarization function
$P^{\rm stat}_Z(t) = \exp(-\gamma_\mu^2 \Delta^2_{Z,{\rm m}}t^2/2) \cos(\gamma_\mu B_0 t)$.
It describes a conventional Gaussian damped spontaneous oscillation. In  
Fig.~\ref{Book_polarisation} is plotted $P^{\rm stat}_Z(t)$ for two values of the ratio
$\Delta_{Z,{\rm m}}/B_0$. From this plot we deduce that this ratio should be smaller than $\approx 1$ to conclude
a spontaneous precession is detected. It is theoretically possible to express the ratio
in terms of the correlation length of the magnetic structure $\xi_{\rm m}$. Assuming a 
ferromagnetic order in a face-centered-cubic compound with lattice parameter $a$ and the muon in 
the octahedral site, we compute $\Delta_{Z,{\rm m}}/B_0$ 
versus $\sqrt{2} \xi_{\rm m}/a$. The result is displayed in Fig.~\ref{Book_ratio}. A relatively short  
$\xi_{\rm m}$ of about ten interatomic
distances is enough to give rise to a detectable spontaneous field. This result is qualitatively
independent of the type of magnetic structure and the muon site. Are ten atomic
distances long enough to call the magnetic structure long-range? In Fig.~\ref{Exp_muon} 
we display a zero-field $\mu$SR spectrum recorded for the triangular system  NiGa$_2$S$_4$.
A spontaneous oscillation is clearly observed at $2.3$~K. Yet, as shown in Fig.~\ref{Exp_neutron} 
the system only exhibits short-range correlations at low temperature. In fact, the lack of 
long-range magnetic order is the main interest of NiGa$_2$S$_4$.

To round up our discussion of the meaning of a damped oscillation for a zero-field $\mu$SR 
asymmetry, we point out that we have recently shown that such a signal may in fact reflect
short-range correlations \cite{Yaouanc13a}. We recall that the Gaussian component field 
distributions as assumed by Kubo and Toyabe explicitly neglect them. 
A simple analytical method to account for them is to extend the component 
field distribution beyond the Gaussian function. This method successfully describes the 
spectrum recorded for the spin-ice system Yb$_2$Ti$_2$O$_7$ at low temperature \cite{Yaouanc13a}.

\section{Commensurate versus incommensurate magnetic order}
\label{incommensurate}
An increasing number of observed spontaneous oscillations are found to be well accounted for 
by a zeroth-order Bessel function of the first kind, rather than a simple cosine. Usually from this observation
it is inferred that the magnetic structure of the compound under study is incommensurate. 
However, the observation of such oscillations does not guarantee the inference to be correct.
To support our argument, we cite two counter examples. A relatively complex yet commensurate structure 
was found in 1992 to yield a field distribution at a nuclear probe reminiscent of the one 
expected for an incommensurate modulated structure \cite{Chiba92}. Recently, the triangular system
Ag$_2$NiO$_2$ previously suggested to be incommensurate using zero-field $\mu$SR data 
was discovered to be commensurate by neutron diffraction \cite{Sugiyama06,Nozaki13}.
While the determination of a magnetic structure solely from $\mu$SR is questionable, it is pertinent 
to test the reality of a neutron-determined structure by zero-field $\mu$SR measurements. This is so 
because of the difficulty sometimes encountered to determine a magnetic structure from neutron 
diffraction data. In fact, a good practice is to record zero-field $\mu$SR and neutron diffraction data 
for the same sample and to extract the magnetic structure from a combined analysis of the data. We stress that
information on $\xi_{\rm m}$ can be derived from 
neutron diffraction and that the field 
distribution at the muon site depends on it. This is further discussed in the next section.

\section{Numerical determination of the field distribution at the muon site}
\label{correlations}

Here we further consider the effect of short-range correlations on the longitudinal polarization 
function $P_Z(t)$. There is {\sl a priori} no reason for limiting the description of the field 
distribution to the kurtosis as done recently in Ref.~\cite{Yaouanc13a}. Higher terms could matter. 
Here we introduce a numerical method 
which does not have this limitation. At the end of this section it is applied to the analysis of 
the aforementioned Yb$_2$Ti$_2$O$_7$ spectrum. 

Taking the vector field distribution at the muon sites, $D_{\rm v}({\bf B}_{\rm loc})$, 
to be the product of the three Cartesian component field distributions,
\begin{eqnarray}
D_{\rm v}({\bf B}_{\rm loc}) {\rm d}^3 {\bf B}_{\rm loc} = 
D_{\rm c}(B^X_{\rm loc})
D_{\rm c}(B^Y_{\rm loc})D_{\rm c}(B^Z_{\rm loc})
{\rm d} B^X_{\rm loc}{\rm d} B^Y_{\rm loc} {\rm d} B^Z_{\rm loc},
\label{eq:distribution1}
\end{eqnarray}
where $D_{\rm c}(B^\alpha_{\rm loc})$ is the component field distribution along the $\alpha $ 
direction, the zero-field static longitudinal polarization function can be written as follows:
\begin{eqnarray} 
P_{Z}^{\rm stat}(t)& = & \int_{-\infty}^{\infty}   \int_{-\infty}^{\infty}   \int_{-\infty}^{\infty} 
 \left\{ \left({B^{Z}_{\rm loc}\over B_{\rm loc}}\right)^2 + 
\left[1-\left({B^{Z}_{\rm loc}\over B_{\rm loc}}\right)^2 \right]
\cos(\omega_\mu t )  \right\} \cr
& \times &  D_{\rm c}(B^X_{\rm loc})  D_{\rm c}(B^Y_{\rm loc}) D_{\rm c}(B^Z_{\rm loc}) 
{\rm d} B^X_{\rm loc} {\rm d} B^Y_{\rm loc}  {\rm d} B^Z_{\rm loc},
\label{eq:asyStatDef}
\end{eqnarray}
with $B^2_{\rm loc} =   (B^X_{\rm loc})^2 + (B^Y_{\rm loc})^2 + (B^Z_{\rm loc})^2$ and 
$\omega_\mu = \gamma_\mu B$ ($\gamma_\mu=2\pi\times 135.54$ MHz/T).

It is interesting to examine the $P_{Z}^{\rm stat}(t)$ expression. Let us replace 
$B^X_{\rm loc}$ by $-B^X_{\rm loc}$ in Eq.~\ref{eq:asyStatDef}. The function $P_{Z}^{\rm stat}(t)$ 
does not change, but now it is expressed in terms of $ D_{\rm c}(-B^X)$ rather than 
$ D_{\rm c}(B^X)$. The same analysis applies for the $B^Y_{\rm loc}$ and $B^Z_{\rm loc}$ variables.
Consequently, $P_{Z}^{\rm stat}(t)$ is conveniently written in terms of the symmetrized
component field distribution 
$ D^{\rm sym}_{\rm c}(B^\alpha_{\rm loc}) = {1\over 2} 
\left[D_{\rm c}(B^\alpha_{\rm loc}) +  D_{\rm c}(-B^\alpha_{\rm loc})\right]$. 

It is easy to extend our scope to the case of a longitudinal-field measurement of the relaxation.
By definition, an external field ${\bf B}_{\rm ext}$ is applied along the $Z$ direction. To account 
for it we only need to substitute $D_{\rm c}(B^Z_{\rm loc})$ by
$D_{\rm c}(B^Z_{\rm loc} - B_{\rm ext})$ in Eq.~\ref{eq:asyStatDef}. Thus, we implicitly assume that the system is not modified by the external field. Then it must be noted that, contrary to the zero-field case, it is in principle possible to resolve an asymmetric component field distribution.

In fact, there is always some magnetic dynamics. Assuming the strong-collision model to be valid,
$P_Z(t)$ can be expressed with an integral equation in terms of the fluctuation
rate $\nu_{\rm c}$ and $P^{\rm stat}_Z(t)$: 
\begin{eqnarray}
P_Z(t) = P_Z^{\rm stat}(t)\exp(-\nu_{\rm c}t)
+ \nu_{\rm c} \int^t_0 P_Z (t - t^\prime) P_Z^{\rm stat} (t^\prime)
 \exp(-\nu_{\rm c} t^\prime)
{\rm d} t^\prime.
\label{eq:dynamics_1}
\end{eqnarray}

To proceed further, as for the analytical analysis of $P_Z(t)$ given in 
Ref.~\cite{Yaouanc13a}, we shall assume the three Cartesian component field distributions to
be identical. Thus we can write 
$P_Z(t) = P_Z(t, \nu_{\rm c}, \left \{D_{\rm c}(B^\alpha_{\rm loc})\right \})$. 
Experimentally, the measured asymmetry is described as the sum of two components:
\begin{eqnarray}
A(t) = a_0 P^{\rm exp}_Z(t) = a_{\rm s} P_Z(t) + a_{\rm bg},
\label{eq:exp}
\end{eqnarray}
where $a_{\rm s}$ and $a_{\rm bg}$ are the initial asymmetry for the sample and background, 
respectively. We stress that experimentally there is always some background, i.e.\ some 
implanted muons miss the sample. Our experience shows that it should always be taken into account 
for a reliable analysis. For the numerical analysis we shall consider $N_B$ discrete field values 
$B^\alpha_{{\rm loc},i}$ in the finite field range $[B_{\rm min}, B_{\rm max}]$. To determine 
$D _{\rm c}(B^\alpha_{\rm loc})$
we need to minimize the following chi-square:
\begin{eqnarray}
\chi^2 = \sum _{i = 1}^{N_{\rm t}} {\left [d(t_i) - A (t_i) \right ]^2 \over \sigma_i^2}.
\label{eq:chi}
\end{eqnarray}
Here $N_{\rm t}$ is the number of discrete equidistant time channels of the measured asymmetry data, 
$d(t_i)$ the measured asymmetry at time $t_i$, and $\sigma_i$ the related standard deviation.
For a reliable fit $\chi^2$ should be roughly equal to the number of degrees of freedom,
i.e.\ $\chi^2  \simeq N_{\rm dof} = N_{\rm t} - N_{\rm p}$, where $N_{\rm p}$ is the number of free 
parameters. 
If we were to solely minimize $\chi^2$ the noise for 
$D_{\rm c}(B^\alpha_{\rm loc})$ would be substantial. To proceed we shall use the maximum entropy 
(ME) concept, and therefore we shall minimize
\begin{eqnarray}
F = \chi^2 -\lambda {\mathcal E},
\label{eq:ME}
\end{eqnarray}
where $\lambda$ is a Lagrange parameter and ${\mathcal E}$ the entropy, i.e.
\begin{eqnarray}
{\mathcal E} = - \sum_{i = 1}^{N_{\rm B}} 
\left [D_{\rm c}(B^\alpha_{{\rm loc},i}) \delta B^\alpha_{{\rm loc},i} \right ]
\ln \left [D_{\rm c}(B^\alpha_{{\rm loc},i}) \delta B^\alpha_{{\rm loc},i} \right ],
\label{eq:entropy}
\end{eqnarray}
where $\delta B^\alpha_{{\rm loc},i}$ is the field distribution step which is taken uniform here.
The basic idea is to minimize $\chi^2$ while maximizing ${\mathcal E}$.

As a first example of the use of the technique described above, we analyzed the Yb$_2$Ti$_2$O$_2$
spectrum already considered with the analytical method in Ref.~\cite{Yaouanc13a}. There it was 
shown that the small applied longitudinal field did not have any effect on the spectrum. So it 
will be analyzed as a zero-field spectrum. We take $B_{\rm max} = -B_{\rm min} = 20$~mT and 
$N_{\rm B}$ = 23 points to describe the field distribution. Since a symmetrized distribution is considered, 
it counts only to 12 field points besides the $N_{\rm p}=3$ parameters $a_{\rm s}$, $a_{\rm bg}$, and $\nu_{\rm c}$. 
Two methods were tested 
for the $F$ minimization: the reverse Monte Carlo (RMC) and the conventional Gauss-Newton (GN) 
methods. We found the first method better in terms of speed of convergence. In addition,
it has the ability to converge to the global minimum, i.e.\ local minima are avoided.
The RMC method is described in the appendix to this report. In Fig.~\ref{ME_distribution} 
\begin{figure}
\begin{minipage}[t]{3in}
\includegraphics[height=2.1in]{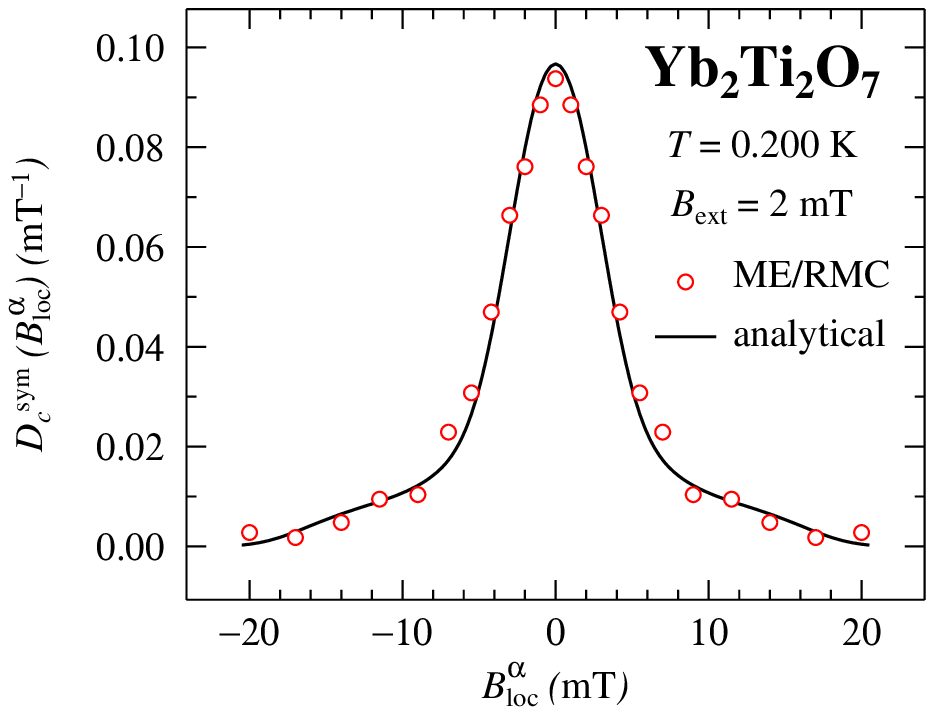}
\caption{Comparison of the results from two methods for the determination of the symmetrized 
component field distribution at the muon site derived from measurements of the longitudinal asymmetry 
for a powder sample of Yb$_2$Ti$_2$O$_7$ \cite{Hodges02} below the temperature at which the specific 
heat displays a sharp peak. While the result from the maximum entropy concept is obtained in this 
work, the analytical analysis was previously presented \cite{Yaouanc13a}.
} 
\label{ME_distribution}
\end{minipage}
\hfill
\begin{minipage}[t]{3in}
\includegraphics[height=2.1in]{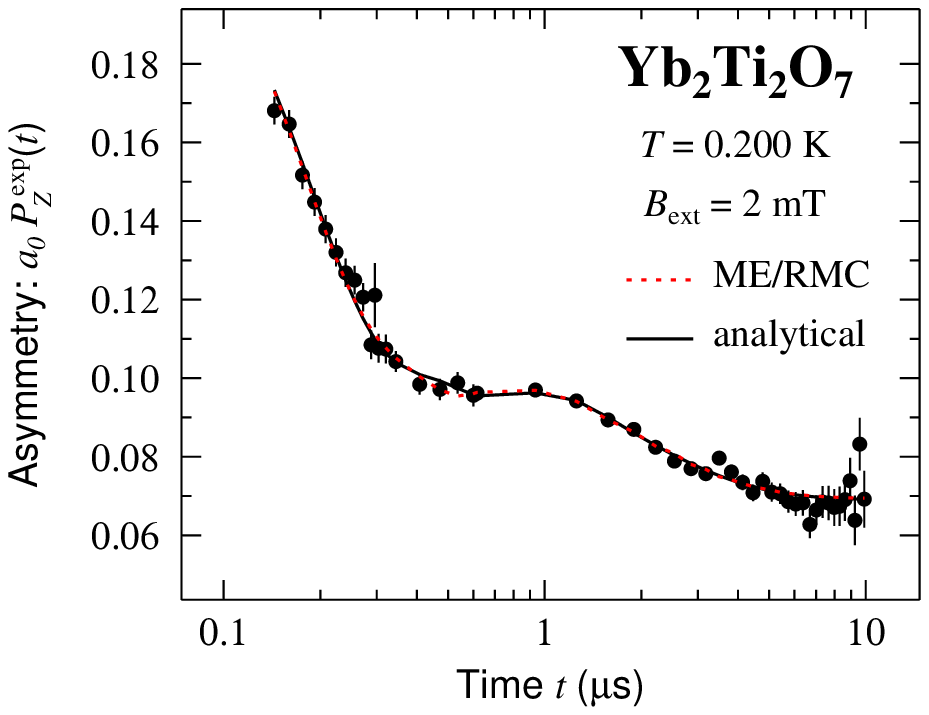}
\caption{The measured longitudinal asymmetry $a_0 P^{\rm exp}_Z(t)$ displayed in a semi-logarithmic 
plot and results from its 
analysis in terms of the two symmetrized field distributions displayed in 
Fig.~\ref{ME_distribution}.
}
\label{ME_relaxation}
\end{minipage}
\end{figure}
we display the symmetrized field distributions extracted from the spectrum shown in 
Fig.~\ref{ME_relaxation}. It is rewarding that the previous analytical result \cite{Yaouanc13a}
is consistent with the ME/RMC data. This provides a justification to limit
the description of the field distribution to the kurtosis in this case. Obviously, the analytical
analysis is performed much faster than the numerical one.

\section{Discussion}
\label{discussion}

The $\mu$SR method is recognized for its ability to easily detect magnetic transitions.
However, one should not overinterpret the experimental data, such as proposing a long-range
magnetic order when the features of the asymmetry does not allow to do it. Inferring a 
magnetic structure to be incommensurate with only $\mu$SR data at hand does not seem reasonable. 
To further complicate the state of affairs, the observation of the expected initial
asymmetry without a spontaneous oscillation is not a proof of the absence of a magnetic order.
The first recognized counter example was published for the ordered spin-ice
Tb$_2$Sn$_2$O$_7$ in Ref.~\cite{Dalmas06}.

The $\mu$SR technique is well adapted for the study of quasi-static magnetic dynamics, i.e.\ with dynamics 
in the micro- and  nanosecond time scale. Such dynamics is seen in exotic magnetic 
materials such as some heavy fermion and frustrated systems. The longitudinal relaxation
is usually found to strongly deviate from the conventional exponential relaxation. The number 
of analytical formulas available for modeling the data is limited. We have
introduced a numerical method to extract from the measured relaxation the field distribution 
at its origin. It supplements a recently proposed analytical method. The numerical method 
combines the best of two worlds: the maximum entropy concept first introduced by Rainford and 
Daniell in the $\mu$SR community \cite{Rainford94} and the reverse Monte Carlo algorithm
of common use for analyzing neutron diffraction data; see Ref.~\cite{Laver08} and
references therein. The numerical method proposed here can probably be extended to the analysis 
of spontaneous oscillations, at least in the case of only one muon magnetic site. The computation
discussed here involves three-dimensional field integrations; see Eq.~\ref{eq:asyStatDef}. If $B_0$ 
is sufficiently large relative to $\Delta_{{\rm Z}. {\rm m}}$, the field integrations get restricted 
to one dimension \cite{Yaouanc11}. Then the number $N_{\rm B}$ of discrete field values 
$B^\alpha_{{\rm loc},i}$ can be increased relative to the case considered here such that
the asymmetry spectrum in a rotating frame is properly described, while keeping the 
computing time reasonable.   

\ack
AY gratefully acknowledges Dr J.~Sugiyama for a useful communication.

\appendix

\section{The reverse Monte Carlo method}
\label{RMC}

We write
$F = F\left(a_0, a_{\rm bg}, \nu_{\rm c}, \left\{ D_{\rm c}(B_{\rm loc,i}^\alpha)\right\}\right) \equiv F({\bf r})$,
i.e.\ the set of parameters for $F$ --- 
$a_0$, $a_{\rm bg}$, $\nu_{\rm c}$,  $\left \{D_{\rm c}(B_{\rm loc, i}^\alpha)\right\}$ ---
is denoted with the single vector ${\bf r}$ of dimension $N_r = N_{\rm p}+ N_{\rm B}$.

Each iteration $j$ of the analysis consists of the following steps (see e.g. \cite{Laver08,McGreevy01}):\\
(1) A component $r_i$ of ${\bf r}$ is randomly chosen ($i= 1,2,...,N_r$).\\
(2) We change $r_i$  by a factor of $(1+{\mathcal S}\epsilon)$, where, in our case, $\epsilon = 0.03$ 
(i.e.\ 3\% of change) with the sign ${\mathcal S}=1$ or ${\mathcal S}=-1$ randomly chosen. Thus, we obtain 
the new vector ${\bf r}_{\rm new}$.\\
(3) $F$ at iteration $j$, that is $F_j$, is computed using Eq. \ref{eq:ME} for 
${\bf r} = {\bf r}_{\rm new}$. \\
(4) If $\Delta F_j \equiv F_j - F_{j-1}<0$ the iteration is successful and ${\bf r}_{\rm new}$ is accepted 
and stored (${\bf r}_{\rm new} \rightarrow {\bf r}$). Otherwise ${\bf r}_{\rm new}$ is rejected and 
previous ${\bf r}$ is kept. The next iteration is repeated from step (1). There is an exception at this 
stage: for $\Delta F_j$ positive there is still a possibility to accept the iteration, and therefore 
${\bf r}_{\rm new}$, with the probability ${\mathcal P} = \exp(-\Delta F_j/p)$. The parameter $p$ is chosen 
such that about half of the iterations are accepted \cite{Laver08}. We found $p = 0.003$ to be a proper 
choice. The $p$ value depends on $\Delta F_j$ and consequently on $\epsilon$. This mechanism of accepting 
the iteration with a positive $\Delta F_j$ turned out to be important for escaping local minima of 
$F({\bf r})$. The larger the number of independent parameters $N_r$ the higher the probability 
to converge to a local minimum.

This algorithm depends on several parameters: $\epsilon$, $p$, and $\lambda$. The larger $\epsilon$ the faster the convergence at the early stage of iterations. However, one should consider that $\epsilon$ determines also the accuracy of the fitted parameters. As a variation of this algorithm one can gradually reduce $\epsilon$ with increasing iteration number $j$. The parameter $\epsilon$ is related to $\Delta F_j$ and consequently to the parameter $p$, determining the probability to accept an iteration with a positive $\Delta F_j$. Obviously the larger $\epsilon$ the larger the absolute value of $\Delta F_j$. In our implementation of the algorithm we analyze the average number of accepted and rejected iterations within the last 10 iterations and we increase or reduce parameter $p$ to reach the 50\% goal for accepted iterations. The value of $\lambda$ determines the weight of entropy in the function $F({\bf r})$. Entropy determines the probability of a distribution, while $\chi^2$ determines the fit quality. $\chi^2$ is a statistical quantity with an expectation value of $N_{\rm dof}$ and standard deviation $\sqrt{N_{\rm dof}}$. Consequently, for a good fit $N_{\rm dof}-\sqrt{N_{\rm dof}}\lesssim \chi^2  \lesssim N_{\rm dof}+\sqrt{N_{\rm dof}}$. Note that the number of free parameters used for the evaluation of $N_{\rm dof}$ is $N_{\rm p}$ i.e.\ the parameters entering the entropy calculation are excluded. If $\chi^2$ is too small one should increase $\lambda$ and conversely for a large $\chi^2$, $\lambda$ should be reduced. The result does not change significantly with a reasonable variation of $\lambda$ for $\chi^2  \simeq N_{\rm dof}$. The particular value of $\lambda$ must be adapted according to the number of field points $N_B$. To ensure the reliability of the result it is important to check that the condition $\chi^2  \simeq N_{\rm dof}$ is satisfied. The value of $N_B$ can be arbitrarily chosen, however one should consider that the computation time increases as $N_B^{3}$. A minimum number of points in the distribution is obviously required. 

In the case of our spectrum, a stable solution was reached after about 500 successful RMC iterations. Its stability was checked by extending the 
computation up to 2000 successful RMC iterations. The computation was started with an unbiased 
distribution. i.e.\ assuming a flat distribution. Note that despite the large number of iterations required
to converge, $F({\bf r})$ is evaluated only once per iteration. Thus the convergence 
rate is quite fast considering the large number of parameters $r_i$.

Obviously, other than RMC methods may be used for a nonlinear minimization such as that of $F$. Note that the linear optimization method 
suggested for transverse field $\mu$SR analysis in Ref. \cite{Rainford94} does not apply in present case. 
The relation between time and field domains given by Eq. \ref{eq:asyStatDef} is highly nonlinear and consequently iterative 
nonlinear minimization algorithms should be used. The obvious merits of RMC are its simplicity, a reasonable computation efficiency, and 
most importantly, possibility to avoid local $\chi^2$ minima in parameter space due to the mechanism of accepting also the iterations 
with reduced fit quality. The presented method can be very useful in analysis of $\mu$SR data of correlated and magnetically ordered 
systems in zero or longitudinal field.

\section*{References}
\bibliographystyle{iopart-num-wo-url.bst}
\bibliography{reference}

\end{document}